\documentclass[10pt]{article}
\usepackage{authblk}
\usepackage{bm}
\begin {document}

\title {\bf {Quantum inertia of non-local matterspace with kinetic and metric counter-stresses}}

\author[a] {Igor {\'E}. Bulyzhenkov}

\affil [a] {\small %Institute of Gravitation and Cosmology,
	 Peoples\rq{} Friendship University of Russia, 

6 Miklukho-Maklaya St., Moscow, 117198, Russia 

 e-mail: ibphys@gmail.com
%Moscow Institute of Physics and Technology,  9 Institutsky lane, Dolgoprudny, 141700, 
    }

\date{}

%%%%%%%%%%%%%%%%% END OF PREAMBLE %%%%%%%%%%%%%%%%

%\begin{document} 


%\baselineskip24pt


\maketitle

%\begin{sciabstract}
% {\bf Abstract}. 
{\small
  New microscopic geodesics  with local balances of kinetic and metric  four-potentials  can maintain the non-local mass mechanism of continuous matterspace due to positive energies of kinetic micro-oscillations. The monistic pan-unity of the inertial continuum of kinetic energy is quantitatively substantiated by the Lomonosov local push on macroscopic densities in  inverse square interactions of visible bodies. The monistic mechanics of inertial fields replaces the dual model of separated particles and their pulling fields or forces at a distance. The positive kinetic pressure of Poincar{\'e} withstands the metric stress for equilibrium densities and prevents the radial collapse of non-local self-organizations, despite the observed fall of probe (extremely dense) densities in rarefied regions of the correlated continuum of the emerging mass-energy. The four-dimensional geometrization with the inherent sub-symmetries for flat matterspace and adaptive kinetic-metric time can be useful for modeling stationary and auto-pulsing astrosystems, as well as for the modeling of vortex hydrodynamics with non-local feedback and micro-macro-mega-entanglement in the monistic field of quantum inertia.  
 }    

%\end{sciabstract}

\bigskip\normalsize


%\bigskip 
%\noindent 
%{\bf PACS}:  {04.20.Cv; 
 %	 	03.70.+k;	 04.50.Kd;02.40.Hw}


\bigskip 
\noindent {\bf Contents}

1. Introduction

1.1 Motivation for Cartesian matter-extension 

1.2 Lomonosov local push and Poincar{\'e} counter-pressure

2. Geodesic relations with microscopic and macroscopic fields

2.1 Kinetic potentials in General Relativity 

2.2 Kinetic-metric balances for microscopic oscillations 

2.3 Probe bodies in the macroscopic geodesics 

2.4 From microscopic geodesics to macroscopic fields

3. Mathematical identities for microscopic balances  

3.1 Euclidean 3-interval and local adaptive time

3.2 Field stresses and inductions

3.3 Adaptive geodesics  and kinetic pan-unity
of the non-local cosmos

4. Kinetic creation of  non-local mass-energy and measurable metric gradients

4.1 Correlated inertial fields and averaged metric replicas 

4.2 Method of kinetic potentials to create measurable mass-energy 

5. Quantum fields of correlated inertia

5.1 Kinetic nature of Poincar{\'e} counter-pressure    

5.2 Quantization of inertial fields and volume integrals of non-local mass-energy

6. Discussion %Conclusion

\section {Introduction }

\subsection {Motivation for Cartesian matter-extension }

The dual theory of relativistic particles and massless field-mediators traditionally tends to use the Dirac delta-operator or the point mass in empty space, referenced by Newtonian mechanics. Such classical dualism for kinetic and gravitational  energies for particles and fields, respectively, cannot resolve the unphysical divergence in the point center of radial energy. The Cartesian matter-extension \cite {Des} could avoid material singularities and energy divergence in the integral distribution of continuous masses or electric charges.  But there is a new challenge here.  The extended particle  must somehow stabilize the Coulomb repulsion or the Newtonian collapse of material densities due to the hypothetical  Poincar{\'e} pressure \cite {Poi}. An inward negative pressure should prevent an elementary charge cloud from disintegrating under its own Coulomb interaction, while an inward positive pressure is required to stabilize  a continuous distribution of any mass-extension. 

The microscopic origin of the  Poincar{\'e} pressure artifact was not satisfactorily discussed in the 4/3 problem for a Thompson non-local charge \cite{BulT}. Lomonosov\rq{s}  arguments in favor of only local pushes \cite{Lom, BulN, BulCol} instead of Newtonian (and Coulombian) distant interactions did not change the university program to a continuous electron with balancing internal pressure.
Below, we consider the mechanical domain for non-local distributions of inertial mass-energy. Here, we recall the Umov\rq{s} ether fraction with the speed-dependent \lq rest-energy\rq{} \cite {Umo} and the  Mie theory of field matter \cite {Mie} to 
relate the macroscopic mass density, the  Lomonosov\rq{s} local push and the Poincar{\'e} counter-pressure to primary microscopic oscillations of the continual matterspace.     

Similar to electrostatics, the inward pressure in a spherical mechanical organization must somehow prevent the gravitational collapse of steady mass densities  
$ m r_o / 4 \pi r^2 (r^\prime + r_o)^2 $, analytically derived \cite {Bul} for the extended particle with the strong, post-Newton  field  $ (- Gm) / r^\prime(r^\prime + r_o), r_o \equiv  mc^2 / \varphi^2_{\!_G}$, $\varphi_{\!_G} \equiv c^2 / {\sqrt G}  = 1.1 \times 10^{22} kg^{1/2}m^{1/2}/s$, in the co-moving system of references, where $r^\prime \equiv |{\bf x}^\prime|$ and $ r^\prime \Rightarrow  r$ for statics. 
The geodesic motion can be analyzed (independent of electricity) due to the well-tested  metric-kinetic relations of Einstein $ \mu^\prime c^2 u^\nu \nabla_\nu (-u_\mu) \equiv \mu^\prime c^2 u^\nu (\nabla_\mu u_\nu - \nabla_\nu u_\mu) =
\mu^\prime c^2 u^\nu (\partial_\mu u_\nu - \partial_\nu u_\mu) = 0 $ for a probe body or a passive mass density $ \mu^\prime = const $ in an external metric field.

\subsection {Lomonosov local push and Poincar{\'e} counter-pressure}

The 1914 geodesics of probe massive points in already established (external) fields serve satisfactorily for many local measurements and observations of small probe bodies. But these local geodesic relations cannot reveal mathematical  details of the non-local organization of external fields. Some non-local laws for a correlated self-organization  of metric fields should clarify their steady shapes around the  center of inertia of the distributed mass-energy $Mc^2=const$. The correlated densities of the isolated  mass-energy and their field stresses should be non-locally organized in some unknown way, either in a line of the Shannon information in kinetic distributions, or in a line of the adaptive Poincar{\'e} counter-pressure. In any case, the nonlocal correlations within geometric space-time require the instantaneous information control and integral preservation in the whole mass-energy, momentum and angular momentum of a closed mechanical system.

The accelerating fields locally represent an external system of distributed mass-energy with its non-local charge integral $ {\sqrt G} M \equiv q_M = const $. Non-equilibrium densities of Lomonosov\rq{s} \lq gravitational\rq{} liquid can move towards  their equilibrium shapes according to the ethereal worldview of Russian Cosmism. The corresponding statement that invisible microscopic motion is everywhere behind the observed mechanical events was made by Plato, Aristotle, Descartes, and many other philosophers. Regardless of what and how one  measures locally, the macroscopic properties of the probe density $ c^2\mu^\prime (x^\prime) $ in the co-moving frame $ \{ x^{\prime o}; x^{\prime i} \}$  can be associated, thanks to Einstein, with the metric four-vectors $g^{\mu\nu}u_\mu = u^\nu \equiv dx^\nu / {\sqrt {g_{\rho\lambda} dx^\rho dx^\lambda }}$. We also employ metric-kinetic four-potentials $U_\mu \equiv \varphi_{\!_G} u_\mu \equiv c^2u_\mu / {\sqrt G}  $ for mechanical charges ${\sqrt G} m$ and  geometrically (non-locally) organized derivatives-divergences $\varphi_{\!_G} f_{\mu\nu} \equiv (\nabla_\mu U_\nu - \nabla_\nu U_\mu) = \varphi_{\!_G} (\partial_\mu u_\nu - \partial_\nu u_\mu)$, $\varphi_{\!_G}f^{\mu\nu} \equiv (\nabla^\mu U^\nu - \nabla^\nu U^\mu)$. For instance, the laboratory mass-energy density $ c^2\mu^\prime  \gamma(x) {\sqrt {g_{oo}(x)}}$ of the moving scalar density $c^2\mu^\prime$ depends on its kinetic potential $\gamma (x)$ (or Lorentz factor) and the metric potential $g_o(x) \equiv{\sqrt {g_{oo}(x)}} $ in the geometric formalism with continuous affine connections-correlations, while the momentum density $c\mu^\prime u_i\equiv c\mu^\prime \gamma (g_i-v_i)$ depends on the three-component kinetic potential $\gamma (-v_i/c)$ and the three-component metric potential $\gamma g_i \equiv \gamma g_{oi}/g_o$.

The lack of microscopic mechanisms for equilibrium balances of emerging metric fields within the spatial self-distribution  of non-local mass-energy prevented the kinetic interpretation of the continuous self-energy $mc^2$ in the cosmic continuum of any elementary organization. 
Therefore, field distributions of quantum (wave) matter still have no physical replica in the Einstein\rq{s} theory of metric fields, which is mainly focused on precise measurements with probe bodies.  By accepting the universal nature for non-local self-organization in both quantum and classical distributions of the kinetic mass-energy integral $mc^2$, Cartesian thinkers today can look  at the adaptive self-organization of correlated densities within the monistic pan-unity of the material cosmos.   Local Lomonosov pushes originate from the microscopic (i.e. ethereal that time) properties of material space and can now be described  in macroscopic terms by four metric potentials $g_\mu \equiv g_{o\mu}/g_o$ from Einstein\rq{s} General Relativity. 

The monistic approach to the kinetic basis of everything can introduce a new force of inertia (or Poincar{\'e} pressure) to balance the Newtonian force or the local Lomonosov push in non-local metric organizations of continuous matterspace. The microscopic oscillations-fluctuations or inward auto-pulsations (like quantum   \lq{} zitterbewegung\rq{)} become the primary cause of emerging macroscopic masses and time-averaged metric fields in macroscopic measurements.
The microscopic self-organization of the purely kinetic (monistic) cosmos jointly resolves  both the Bentley gravitational paradox and the Poincar{\'e} pressure hypothesis. The limiting factor for Cartesian matter-extension and its monistic pan-unity in Russian Cosmism was and is that the monistic worldview conceptually contradicts the empty-space doctrine for macroscopic fields between localized bodies.

\section {Geodesic relations with microscopic and macroscopic fields
}

\subsection {Kinetic potentials in General Relativity 
}

The metric fields of the Earth (or the Sun, etc.) locally represent  the inertial charge integral, $ {\sqrt G} M_{Earth} \equiv 
q_{_M} > 0$, distributed around the  center of spherical symmetry. Stationary or pulsating self-assembly of  metric fields can occur for any  quasi-equilibrium distribution in the world hierarchy of quasi-isolated energies. In other words, non-equilibrium  matterspace-time sub-organizations can cyclically change their material densities around equilibrium shapes in the case of negligible dissipation and in the absence of structural bifurcations. 

Inelastic (wave) energy exchange with other sub-organizations slowly dumps or pumps such geodesic auto-pulsations of continuous densities. Moreover, auto-pulsating densities can evolve into new  macroscopic or megascopic evolutionary shapes due to inward 
wave exchanges under the constant mass-energy integral $ {\sqrt G} M \varphi_{\!_G} \equiv Mc^2 = const$ after collisions with other  (almost completely isolated) systems. Again,  
 an inward dissipation (or inward wave exchange) for metric shaping of the elementary system keeps its inertial charge  $q_M \equiv Mc^2 /\varphi_{\!_G} = const$. External forces or  radiation disturb the metric self-organization process and can change its mass-energy integral, as well as the affine connections of the field distribution.    

The probe mass density $ \mu_p^\prime (x^\prime) = const $ is to be defined in General Relativity in the co-moving frame $ \{ x^{\prime o}; x^{\prime i} \}$, but not in the laboratory one $\{ x^{o}; x^{i} \}$. Einstein described the free fall of this passive density or the point mass scalar in external fields  by the adaptive four-velocity $u_\mu \equiv g_{\mu\nu}u^\nu \equiv g_{\mu\nu}dx^\nu / {\sqrt {g_{\rho\lambda} dx^\rho dx^\lambda }}$, with $u_\mu u^\mu \equiv 1$. We similarly define    
 the metric-kinetic four-vector potential $U_\mu \equiv \varphi_{\!_G} u_\mu \equiv c^2u_\mu / {\sqrt G} \equiv g_{\mu\nu}U^\nu $ and employ the anti-symmetric field $\varphi_{\!_G}f_{\mu\nu} \equiv (\nabla_\mu U_\nu - \nabla_\nu U_\mu) = \varphi_{\!_G} (\partial_\mu u_\nu - \partial_\nu u_\mu)$, with $\varphi_{\!_G}f^{\mu\nu} \equiv (\nabla^\mu U^\nu - \nabla^\nu U^\mu)
\equiv \varphi_{\!_G}g^{\mu\alpha} g^{\nu \beta}f_{\alpha\beta}$.  Recall that the laboratory energy density $\mu_p(x)c^2 = \gamma(x) c^2\mu_p^\prime (x^\prime) $ depends on its kinematic potential $\gamma = 1/ (1-\beta_i\beta^i)^{1/2}$ of probe bodies in Special Relativity (SR). 

In General Relativity (GR), the SR kinetic energy was also made  path-variable  by the metric potential $g_o(x) \equiv {\sqrt {g_{oo} (x)}}$ through the monistic product in the unified GR potential $g_o c^2\gamma$ for scalar masses,  
$ \mu_p^\prime (x^\prime) \gamma(x)  {\sqrt {g_{oo} (x)}}c^2 > 0 $. The unified kinetic product in GR does not operate with a sum of two  energies as in Newton\rq{s} dual model for independent kinetic and gravitational notions. We emphasize  the purely kinetic, monistic meaning of the total relativistic energy by the conceptual multiplication $\gamma g_o$ of positive kinetic, $\varphi_{\!_G}\gamma >0 $, and positive metric, ${\sqrt {g_{oo}}} \equiv 1 + (\varphi_{gr}/c^2) > 0$, potentials for the inertial density $c^2\mu^\prime$. Again, General Relativity monistically admits only positive densities of kinetic inertia, ${\sqrt G}\mu_p^\prime U_o \equiv \mu_p^\prime \gamma (c^2 + \varphi_{gr}) \geq 0 $,  rather than the dualistic sum of different  notions: the kinetic energy density $\mu^\prime \gamma c^2$ of the massive particle and the negative gravitational contribution, $\mu^\prime \gamma \varphi_{gr} < 0$, from \lq inertial-free\rq{} fields. 

The monistic approach of  Russian cosmists  to purely kinetic  densities of the ambient matterspace with etheric and thermal contributions to its measurable inertia, is conceptually consistent with the monistic reading of positive mass-energy in General Relativity, but not with the Newtonian duality of localized masses and massless fields/forces. In other words, the monistic mechanics of massive, inertial fields  requires for the non-dual reading \cite {EI} of mass-energy in General Relativity and conceptually rejects the  Standard Model of inertial particles and field mediators in empty space of the classical mechanics. 
Particle\rq{s} mass-energy and field\rq{s} mass-energy are the same notion in the physical reality as first claimed in the Platonic  dialogues: \lq\lq matter and space are the same\rq\rq{.} 
 
Below, emphasizing the primacy of available observations and local measurements in 3D space, we traditionally relate the geodesic acceleration of probe bodies to the metric stresses of macroscopic fields. And we relate the true origin of these fields in macroscopic measurements  to the immeasurably fast  vibrations-oscillations of the Lomonosov-Umov microscopic ether or the continuous matter-extension of Descartes. The microscopic kinetics, missing in the phenomenological models of dual energies, explain the mass creation mechanism of macroscopic reality. The microscopic kinetics of continuous matterspace also explains the inward force of inertia (or Poincar{\'e} stabilizing counter-pressure), also missing in dual macro-mechanics with the gravitational collapse of separated masses.

The ultra-fast microscopic oscillations enable a monistic induction of both measurable masses and measurable metric stresses.
 New relativistic geodesics should be employed  to derive the time-averaged metric fields for macroscopic measurements in always flat 3-space.   New microscopic relations form correlated metric fields for the macroscopic dynamics of probe bodies. Very high, visible densities of the probe body are not in equilibrium with the self-organized kinetic densities of the radial Earth beyond its  atmosphere. And Lomonosov\rq{s} pushes  (or Einstein\rq{s} metric stresses) locally accelerate these densest areas of probe bodies according to the 1914 geodesic relations with macroscopic metric fields. 
The latter do not repeat the ultra-fast oscillations-fluctuations of microscopic fields, but generate the time-averaged gradients of the emerging metric potentials in the nonlocal matterspace with adaptive time rates.

\subsection {  Kinetic-metric balances for microscopic oscillations 
	      }

Classical textbooks propose to relate the covariant  acceleration, $ m_p c^2 D(-u_\mu)/Ds$ $ = F_{\mu\nu}J_e^\nu/c$
\cite {Lan},  of the point mass $m_p$ with its electric charge $(-e_p)$ in external electromagnetic fields $F_{\mu\nu} \equiv \nabla_\mu A_\nu - \nabla_\nu A_\mu$ with the local Lorentz force $ F_{\mu\nu}J_e^\nu/c$
based on the translation four-vector $J_e^\nu = (-e_p)c u^\nu$. But where are self-forces for such a macroscopic current?
The formal contraction $J_m^\nu f_{\mu\nu}/c = 0$ of the probe mass 4-vector $J_m = cm_p u^\mu $ to the similar mechanical tensor $f_{\mu\nu} \equiv \nabla_\mu u_\nu - \nabla_\nu u_\mu$ with external metric fields duplicate the Einstein geodesics of 1914,
 $m_p c^2 D (-u_\mu)/Ds \equiv  m_pc^2 u^\nu \nabla_\nu (-u_\mu)  = m_pc^2u^\nu f_{\mu\nu}  = 0$,  due to the metric-kinetic qualities $u^\nu \nabla_\mu  u_\nu$ $ \equiv \nabla_\mu (u^\nu u_\nu) /2 \equiv 0$.
Again, there is no self-action here because of the formal  linearity  $c^2 J_m^\nu \propto  \mu_p^\prime c^3 u^\mu$ for the mechanical current density $J_m^\nu$ and the coordinate translations of time-averaged  probe densities $\mu_p^\prime c^2$.

It is important to  emphasize  that local measurements of space-time gradients of metric potentials $g_\mu \equiv g_{o\mu}/ {\sqrt {g_{oo}}}$ by passive (probe) bodies and the correlated self-organization of these metric fields  before measurements are different physical processes. One should not  identify the divergence-free current density $j^\nu (x)\equiv - c\varphi_{\!_G} \nabla_\lambda f^{\lambda\nu}(x)/4\pi $ for the adaptive self-assembly of distributed mass-energy with the Newton-Euler flow ${\sqrt G} \mu_p^\prime c u^\nu (x) $ of the passive density  ${\sqrt G} \mu_p^\prime = const$. Metrically organized densities of material space within the volume integral of the non-local mass-energy $\varphi_{\!_G}\int {\sqrt {|g_{ij}|}} {\sqrt G} \mu^\prime (x^\prime)d^3  x^\prime =  mc^2 \equiv r_o \varphi^2_{\!_G} = const$
do not obey  Newtonian mechanics of point or volumetric  masses in external fields or Euler/Navier-Stokes hydrodynamics with forced accelerations without local self-actions or adaptive feedback \cite{BulTur}.

According to the mathematical theorems of  G$\rm{\ddot o}$del there are no completely isolated or purely metric organizations of \lq\lq almost isolated\rq\rq{} systems in the hierarchical structure of kinetic densities. Therefore,  oscillating geometric fields, locally induced by microscopic oscillations, are inevitably   disturbed by external (non-metric) force densities $F^{ext}_\mu$:
\begin {equation} 
   \frac {\varphi_{\!_G}}{c} {f_{\mu\nu}(x)  } j^\nu (x)  \equiv  \frac { \varphi^2_{\!_G}} {4\pi{\sqrt {-g} } }(\nabla_\mu u_\nu - \nabla_\nu u_\mu)
	\partial_\lambda [{\sqrt {-g} } g^{\alpha \nu} g^{\beta \lambda} (\nabla_\alpha u_\beta - \nabla_\beta u_\alpha)]   =  F^{ext}_\mu. 
\end {equation}
Here, one can use $F^{ext}_\mu \Rightarrow 0$ only for quasi-isolated self-distributions with geometric space+time connections and adaptive time rates from the strict microscopic equality  $(cu_\mu)\cdot (cu^\mu) \equiv c^2$. Such a microscopic sub-governance  at adaptive time should result in Euclidean macroscopic space with non-locally correlated connections in order to keep the energy and momentum integrals at all running values of the world coordinate $x^o$. In contrast to the kinetic (primary, reason) - metric (secondary, consequence) relations in microscopy (1), the inverse metric (reason) - kinetic (consequence) relations $\mu_pc^2 D u_\mu/D s $ $\equiv $ 
$\mu_p c^2 u^\nu \nabla_\nu  u_\mu \equiv \mu_pc^2 u^\nu (\nabla_\nu u_\mu - \nabla_\mu u_\nu) \equiv
 \mu_p a^{mk}_\mu = 0$ in General Relativity dictate visible accelerations of the passive macroscopic density $\mu_pc^2 = const$ in a non-locally formed distribution of external fields.

 The isolated geodesics in (1) at $F^{ext}_\mu = 0$ mean long-range correlations of unperturbed micro-oscillations  with the local conservation $\nabla_\mu j^\mu \equiv 0$ everywhere under the geometric organization of non-local mass-energy continuum with the adaptive time $d\tau \equiv g_\mu dx^\mu /c$ for the local current of macroscopic densities. Such a non-local approach to correlated currents and stresses in purely kinetic (monistic) mass-energy of continuous matterspace does not correspond to the dual division of classical energies into kinetic (positive) and gravitational (negative) energies. New geodesics of self-governed ether velocities in (1) recognize the Chinese Yin-Yang dialectics for local counter-balances of kinetic (inertial, yang) stresses and induced metric (yin) replicas within the  correlated continuum of non-local mass-energy. Its volume integral  is constant at all moments of the world parameter $x^o/c$, while the local conservation of mass-energy flows $\nabla_\mu j^\mu =0$ is maintained by the 4D geometric  self-assembly  with  adaptively changing densities in Euclidean 3-section. Such an adaptive self-assembly of pseudo-Riemannian manifolds always  keeps  inherent geometrical symmetries $g_ig_j - g_{ij} = \delta_{ij}$ for the spatial transport of inertial densities with the 3-momentum conservation (confirmed in practice).  The correlated self-assembly of the flat continuum by geometrically organized matterspace + adaptive time reinforces the Mie theory of continual  matter \cite {Mie} and the Einstein-Infeld directive \cite {EI} to non-dual physics of pure field masses.

\subsection {	
 Probe bodies in the macroscopic geodesics 
	    }

Consider GR\rq{s} geodesics for a  macroscopic probe body with the rest mass $m_p = const $ or a scalar mass density $\mu_p^\prime = \mu /\gamma g_o = const$. Coordinate displacements  of probe masses  in external metric fields can be modeled in practice by the translational four-current  ${\sqrt G}\mu_p^\prime c u^\nu (x)$. The 1914 macroscopic equation $ \mu_p^\prime c^2 u^\nu [\partial_\nu (-u_\mu) - \partial_\mu (-u_\nu)] \equiv -\mu_p^\prime a^{mk}_\mu = 0$ describes the geodesic force for mandatory accelerations of the probe charge density  ${\sqrt G}\mu_p^\prime = const$  in external fields: 
\begin {eqnarray}
\cases {
 \mu_p^\prime c^2\gamma  \beta^j [\partial_j (-\gamma {\sqrt {g_{oo}}}) - \partial_o (\gamma \beta_j-\gamma g_j)] \equiv {\sqrt G}\mu_p^\prime \gamma g_o {\bm \beta }\cdot  {\bm E}_{mk} = 0, \cr  
$where$ \ {\bm E}_{mk} \equiv  - (c^2/{\sqrt G}g_o) [{\bm \partial} (\gamma \!{\sqrt {g_{oo}}}) + \partial_o \gamma ( {\bm \beta}- {\bm g})]
             , 
						\cr $and$ \
\mu_p^\prime c^2 \! u^o  [\partial_i (-\gamma \!{\sqrt {g_{oo}}}) - \partial_o (\gamma \beta_j-\gamma g_j)]
\cr  
+ \mu_p^\prime c^2\gamma  \beta^j
[\partial_i  (\gamma \beta_j -\gamma g_i )  - \partial_j (\gamma\beta_i -\gamma g_i )] = 0 \cr
 $or$\ {\sqrt G}\mu_p^\prime (u^o g_o{\bm E}_{mk}   + \gamma  {\bm\beta}  \times {\bm B}_{mk}) = 0,  \cr%\cr
 $where$  \  {\bm B}_{mk}\equiv    \varphi_{\!_G} curl (\gamma{\bm \beta} -\gamma {\bm g}), \   div {\bm B}_{mk} = 0, \cr
 \partial_o  {\bm B}_{mk}  \equiv  - curl ( g^o{\bm E}_{mk}) =
 curl  [ 
(d{\bm x}/dx^o )\! \times \!{\bm B}_{mk}]. 
 }
\end{eqnarray}

Such relativistic generalizations of  Bernoulli flows, ${\bm \beta }\cdot  {\bm E}_{mk} = 0$, and Euler liquids without a self-acceleration feedback, 
${\bm E}_{mk} = -( \gamma  {\bm\beta} /g_ou^o )\times {\bm B}_{mk} $, describe the geodesic dynamics  of probe (macroscopic) masses or non-adaptive probe densities in the absence of non-metric forces. Here, there are no Poincar{\'e} counter-balances in the geodesic motion of passive macroscopic bodies.   Non-metric stabilization or pressure gradients are required by Euler/Navier-Stokes macro-fluids for their static states  ($dx^i / x^o = 0, v_i =c\beta_i = 0, \gamma = 1$) in already formed macro-potentials
 $c^2 g_\mu/{\sqrt G}$ when 
$({\bm \partial}  g_o - \partial_o {\bm g})/g_o \neq 0$. In the absence of stabilizing pressure or other local (external) influences, probe elements always gain the physical 3-velocity $v_i \equiv dx^i/d\tau \neq 0$ due to the 3-acceleration 
in the metric-kinetic balances 
$u^\nu (\partial_i u_\nu - \partial_i u_\nu) = 0$ of 1914. Therefore, an inward pressure should be assigned to stabilize the external geodesic forces in classical liquids. 

The \lq gravitational liquid\rq{} of Lomonosov is self-organized in the whole volume of any mechanical system. In addition to the static, luminous ether of  Huygens, Lomonosov assigned  to his superpenetrating liquid the \lq gyratory motions\rq{} and \lq jitter\rq{} in order to discuss   local pushes of observable probe bodies. Such ethereal rotations and vibrations should be long-range correlated to  maintain the kinetic energy organization on macro- and 
mega-scales. The monistic pan-unity of world material densities is a historical milestone of  Russian Cosmism. It  required monism of living and inert matter (Tsiolkovsky\rq{s} \lq\lq citizen of the Universe\rq\rq{}), as well as non-Newtonian mechanical ideas to describe the kinetic holism with only positive (monistic) mass-energy.

\subsection { From microscopic geodesics to macroscopic fields 
 }

Zeroing the metric-kinetic acceleration $a^{mk}_i$ of the probe mass in the macroscopic balance (2)
does not mean that there are no measurable accelerations and velocities of mechanical bodies (e.g. in the reference frames of distant observers). By accompanying probe bodies, observers can \lq feel\rq{} external metric fields only during drag collisions or due to external pressure with non-metric  forces. Consequently, the motionless ground observer feels only the metric field of the Earth, but not that of the Sun or the Galaxy. Being on a geodesic trajectory (2) near a non-radiating star, any \lq point observer\rq{} in a closed box can measure neither local velocity with acceleration with respect to  the star nor its metric fields. Therefore,  balanced metric-kinetic stresses around such an invisible star cannot be measured by one local observer  in the geodetically accelerated   laboratory. 
The radiation of the Sun reveals the kinematic center of attraction in the solar system. And telescope observations for planetary dynamics could also reveal this center of kinetic-metric counter-stresses in the elastic cosmos around the Sun\rq{s} visible disk. 
How can kinetic-metric balances (1) and averaged metric-kinetic balances (2) bring the invisible matter-liquid of  Lomonosov back into the modern physics of relativistic energies?          

The key question is how the metric organization of matterspace  around a gravitational center can hold the stable macroscopic  densities of averaged mass-energy  in order to maintain the static macro-potential
  ${\sqrt {g_{oo}(r)}}  \approx  
1 - GM/r c^2$ (with $GM_{_E}/R_E c^2 \approx 10^{-9} $ on the visible surface of the Earth)? The local gradients of this macroscopic potential tend to push the inertial bodies one towards each other. And the gravitational collapse of the cosmic masses seems inevitable.  However, this cosmological paradox of Bentley only works for the dualistic model of passive particles in active external fields, but not for the monistic material whole with correlated auto-pulsations or  oscillations within non-locally organized densities in the constant mass-energy integral.  

Below, it is proposed how the Bentley\rq{s} paradox can be solved by repulsive gradients of the averaged kinetic potential  $c^2\gamma (r) $ $= $ $c^2 <1/ {\sqrt {1 -\beta^2}}>_o$  of the non-local mass-energy $M c^2 = const$ with the steady kinetic-metric balance  $\partial_i <[c^2\gamma(r) g_o(r)]>_o = 0 $  corresponding to the Lomonosov gravitational push and the Poincare kinetic counter-pressure. The time-averaged (macroscopic) potential $c^2g_o(r) = c^2 <g_o>_o$ is induced  as a (yin-)replica on very fast (yang-)pulsations of ether velocities. These microscopic (yang-)pulsations are a primary cause of the macroscopic mass densities, the  Lomonosov local pushes, and the Poincar{\'e} stabilizing forces. At the same time, such equilibrium  details about external fields with metric and inertial counter-balances  are not essential for extremely  dense probe masses under their non-equilibrium accelerations in the macro-geodesics (2).

 One can assume highly elastic properties of the microscopic auto-oscillations, because they allow very slow  transitions to  quasi-equilibrium macroscopic shapes during laboratory observation times. These inward oscillations-pulsations within any non-local mass-energy do not contradict to a slow evolution of its cosmic shape in weak external fields. The astronomical observations are always based on the slow dynamics of very dense (visible, probe) bodies which are moving according to the macroscopic relations  (2) with  the time-averaged microscopic four-potential $ <g^{micro}_\mu(x)>_o \equiv g^{macro}_\mu(x)$.

The cyclic self-organization  in the world hierarchy
 of kinetic and metric counter-stresses in (1) leads to  Kant\rq{s} quasi-stable cosmology of visible bodies. And the macroscopic geodesics (2) of probe bodies do not contradict the Nietzsche\rq{s} \lq eternal return of the same\rq{} \cite{Nie}. In fact, Einstein\rq{s} metric theory  quantitatively  supports \cite{BulA} this geodesic return of probe bodies along degenerate Keplerian orbits or vertical falls and rebounds instead of the inevitable  gravitational collapse in the dual mechanics.

\section {Mathematical identities for microscopic balances  
	 }
 
\subsection {Euclidean 3-interval and local adaptive time}

 It is quite difficult to find mathematical solutions of (1) with respect to $g_{\mu\nu}$ at $F^{ext}_\mu = 0$ even for the symmetric affine connections $\Gamma_{\mu\nu}^\lambda = \Gamma_{\nu\mu}^\lambda$ in $\nabla_\mu u_\nu \equiv \partial_\mu u_\nu - {\Gamma_{\mu\nu}^\lambda u_\lambda } $ (when $\nabla_\mu u_\nu - \nabla_\nu u_\mu  = \partial_\mu u_\nu - \partial_\nu u_\mu$). Fortunately for the non-local physics of metric distributions,  high concentrations of adaptive mass-energy correspond to the Euclidean 3-section in the pseudo-Riemann 4-interval, $ds^2 \equiv (c d\tau/ \gamma)^2  \equiv [({g_{oo}} dx^o + g_{oi} dx^i )^2/g_{oo}]   - [(g_{oi} g_{oj}/g_{oo}) - g_{ij}]dx^idx^i $ $
= c^2 d^2\tau  (1 - \delta_{ij}dx^idx^j / c^2 d^2\tau )$, due to six inherent symmetries \cite {Bul}  
$g_{ij} = g_ig_j - \delta_{ij}$
to produce four independent potentials  $g_\mu \equiv g_{o\mu} / g_o $ from
 ten components of the metric tensor $g_{\mu\nu}$. The local ratio $c d\tau/g_odx^o \equiv \dot \tau \equiv 1  + g_i \beta^i {\dot \tau} \equiv 1  / (1-g_i\beta^i)$ for the geodesic motion  
is adaptively adjusted to preserve these inherent symmetries of 4D geometry with the spatial flatness everywhere in the inhomogeneous continuum of non-local matterspace.  
In fact, Euclidean 3-space for extended masses has not been falsified by any of the relativistic tests  \cite {BulTests}. 

Again, ten tensor components $g_{\mu\nu} = g_{\nu\mu}$ with six geometrical bounds $g_{i\mu} = 
g_ig_\mu - \delta_{i\mu} $ (due to the identities $g_{o\mu} \equiv g_\mu g_o$ and $\delta_{io} \equiv 0$) are represented  in the Einstein\rq{s} geodesic relations by four metric potentials $g_\mu(x) \equiv g_{o\mu} / {\sqrt {g_{oo}}} \equiv g_{o\mu}/g_o$, induced by the microscopic kinetics with ethereal velocities.  
The physical time ${d\tau }\equiv g_\mu dx^\mu/c \equiv  g_o dx^o /c (1-g_i\beta^i)$ of correlated flows in non-local mass-energy with the metrically organized space-time is different, for example, from the universal metric time ${d\tau } = g_o dx^o/c $ of a local  observer with $\beta_{obs}^i = 0$. 
In the absence of external forces or non-metric accelerations, one can talk about the adaptive self-organization of pseudo-Riemannian space-time with 3D material continuum, which conceptually preserved Euclidean sub-geometry, $dl^2 \equiv (g_ig_j - g_{ij})dx^idx^j  =  \delta_{ij} dx^idx^j$, for any flows of continuous mass-energy. Such 4D organization of correlated changes maintains the following mathematical definitions in the flat-space approach to non-local matterspace:
\begin {eqnarray}
 \cases {
%\begin {cases} 
g_\mu \!\equiv g_{o\mu}/\!{\sqrt {g_{oo}}}, g_{oo}\! \equiv g^2_o, g_{oi}\! \equiv g_og_i, g_{ij} \equiv g_ig_j \!-\! \delta_{ij}, g_i \equiv \delta_{ij}g^j, g^{ij} \!\equiv - \delta^{ij}, 
\cr
g^{oo}\equiv (1-g_ig^i)/g^2_o, g^i \equiv g_og^{oi} \equiv \delta^{ij}g_j, {\sqrt {-g}} \equiv  {\sqrt {g_{oo} |g_ig_j-g_{ij}|
}} \equiv g_o, 
 \cr 
ds^2 \equiv g_{\mu\nu}dx^\mu dx^\nu \equiv (g_{\mu} dx^\mu)^2 - (g_ig_j-g_{ij})dx^idx^j \equiv c^2 d\tau^2 - dx_j dx^j\cr \equiv c^2d\tau^2
(1-\beta^2)\equiv (g_odx^o/ 1-g_i\beta^i )^2(1-\beta^2) \equiv ( {dx^o}/{u^o})^2, \cr \gamma \equiv 1/{\sqrt {1 - \beta_i \beta^i}}  ,
\beta_i \equiv v_i/c \equiv  (g_ig_j-g_{ij})dx^j/c d\tau \equiv \delta_{ij}\beta^{j},  \cr  
  u_i \equiv \gamma (g_i - \beta_i) ,  u^i \equiv \beta^i \gamma,   u_o \equiv \gamma g_o , u^o \equiv(1-g_i\beta^i)\gamma/g_o  .
}
\end {eqnarray}

\subsection { Field stresses and inductions}

Now we use the geometric identities (3) 
in four kinetic-metric balances (1),  when $F^{ext}_\mu \rightarrow  0 $ and $j_{km}^\nu f_{\mu\nu}/c = 0$, to analyze the non-local self-organization of the Lomonosov matter-liquid.  The divergence-free  distribution of the local  four-currents $j_g^{\mu}(x) \equiv c \varphi_{\!_G} \nabla_\nu f^{\mu \nu}(x)/4\pi $ in the Euclidean 3-space with $ {\sqrt {-g}} = {\sqrt {g_{oo}}}$ at all space-time points means that is
the correlated organization of the pseudo-Riemann manifold due to metrically ordered connections between neighboring  inertial densities. Such a  correlated organization of continuous matterspace under the adaptive local time ${d\tau}\equiv g_\mu dx^\mu/c  $ for moving densities is preserved in (1) by the following microscopic identities for kinetic-metric  (km, not metric-kinetic or mk, as in the 1914 macroscopic geodesics) relations:  
\begin {eqnarray}
 \cases {
 f_{\mu\nu} \equiv \partial_\mu u_\nu - \partial_\nu u_\mu, f_{oo} \equiv 0, f_{oi} \equiv - 
\partial_o  (\gamma\beta_i - \gamma g_i) -  \partial_i (\gamma g_o), \cr
E^{km}_i \equiv \delta_{ij}E^j_{km} \equiv \varphi_{\!_G} f_{oi}/g_o \equiv -\varphi_{\!_G} f_{io}/g_o,  
\cr
0\equiv \varphi_{\!_G} e^{lij}  (\partial_i f_{oj}\! - \partial_j f_{oi} \!- \partial_o f_{ij}\!)/2g_o \equiv   [(curl g_o {\bm E}_{km})^l + (\partial_o  B^l_{km})]/g_o , \cr $or$ 
\ curl  {\bm E}_{km} + (\partial_o {\bm B}_{km}/g_o ) + 
{\bm E}_{km}\times(- {\bm \partial } ln g_o )  \equiv 0   ,  
 \cr  
\{curl g_o {\bm E}_{km} \}^l \equiv e^{lij} [\partial_i (g_o E_j) - \partial_j (g_o E_i)] /2, 
\cr B^l_{km} \equiv - \varphi_{\!_G}e^{lij}(\partial_i u_j - \partial_j u_i)/2, \
{\bm B}_{km} \equiv - \varphi_{\!_G} 
curl [({\bm g} - {\bm \beta }) \gamma], 
\cr
\varphi_{\!_G}j_{km}^\nu\! f_{o\nu}\! \equiv\! \varphi_{\!_G} j_{km}^i\! f_{oi} =0, \ {\bm j}_{km}\cdot {\bm E}_{km} = 0, \  div {\bm B}_{km} \equiv 0, 
 \cr
j_{km}^\nu f_{i\nu} \equiv j_{km}^of_{io} + j_{km}^k f_{ik} \equiv - (g_o j_{km}^oE^{km}_i   +
 [{\bm j}_{km}\times {{\bm B}_{km}}]_i)/\varphi_{\!_G} = 0, \cr   
  g_oj_{km}^o{\bm E}_{km} \cdot {\bm B}_{km} = 0, \ g_oj_{km}^o{\bm E}_{km} \cdot {\bm j}_{km} = 0,
\cr   \partial_o {\bm B}_{km} = curl [({\bm j}/j^o)_{km} \times {\bm B}_{km}  ], 
  \cr\cr
	{4\pi{\sqrt {-g} } } j_{km}^\nu /c  \equiv  { \varphi_{\!_G}} 
	\partial_\lambda [{\sqrt {-g} } g^{\alpha \nu} g^{\beta \lambda} (\nabla_\alpha u_\beta - \nabla_\beta u_\alpha)]
		\cr
 4\pi {\sqrt {g_{oo}}} j_{km}^o / c \equiv \varphi_{\!_G}\partial_\nu {\sqrt {g_{oo}}} f^{o\nu}
 \equiv \varphi_{\!_G}\partial_j g_o f^{oj} \equiv  
 \varphi_{\!_G}\partial_j [g_o (g^{o\lambda}g^{j\nu} f_{\lambda\nu})]   
%- \varphi_{\!_G}\delta^{jk}   \partial_j ( g^{oo}{\sqrt {g_{oo}}}{f_{ok}})
 \cr 
\equiv -  \partial_j \{[ g_o g^{oo}\delta^{jk} + (g^jg^k/g_o) ] g_o E^{km}_k + \varphi_{\!_G}\delta^{jl}g^k f_{kl} \} \equiv                                          - \partial_j D_{km}^j    %  = - {\bm \partial} [{\bm D}_m  + {\bm G}\times {\bm B}_m ],
 \cr
4\pi {\sqrt {g_{oo}}} j_{km}^i / c \equiv   \varphi_{\!_G} \partial_o ({\sqrt {g_{oo}}} g^{i\lambda}g^{o\nu} f_{\lambda\nu}) + 
 \varphi_{\!_G}\partial_k ({\sqrt {g_{oo}}} g^{i\lambda}g^{k\nu} f_{\lambda\nu})  \cr
%=   \partial_o D_m^i - \delta^{is} \varphi_{\!_Q} \partial^p {\sqrt {g_{oo}}}  (\partial_s \gamma \beta_p - \partial_p \gamma \beta_s %) =
\equiv  \partial_o D_{km}^i  - [curl  (g_o{\bm H}_{km}) ]^i  ,%\ H^m_k \equiv - e_{kij} H_m^{ij}/2,
\cr\cr
%{{B}^k_m} \equiv  - e^{kij}B^m_{ij}/2 \equiv \varphi_{\!_G}  [curl ({ \bm \beta}-{\bm g})\gamma ]^k,  B^m_{ij} \equiv  \varphi_{\!_G}f_{ij}  , 
%H_m^{ij} \equiv {\sqrt {g_{oo}}} f^{ij}\varphi_{\!_G},
%{\bm H}_m\equiv {\sqrt {g_{oo}}} {\bm B}_m,
D_{km}^i \equiv - \varphi_{\!_G}{\sqrt {g_{oo}}} f^{oi} \equiv   [(1- g_jg^j)\delta^{ik} + g^ig^k ] (\varphi_{\!_G}f_{ok}/g_o) + \varphi_{\!_G}\delta^{il}g^k f_{kl}, \cr
{\bm D}_{km} \equiv  {\bm E}_{km}  +{\bm g}\times [{\bm g}\times {\bm E}_{km} +  {\bm B}_{km} ] 
\equiv {\bm E}_{km}  + ({\bm g}\times {\bm H}_{km}), 
\cr   H^i_{km} \equiv   B^i_{km} + [{\bm g}\times {\bm E}_{km}]^i,   \  g_oj^o{\bm E}_{km} = {\bm B}_{km}\times  {\bm j},
\cr 
[curl g_o({\bm B}_{km} + {\bm g}\times {\bm E}_{km})]^i \equiv
[curl (g_o{\bm H}_{km})]^i \equiv - \varphi_{\!_G}\partial_k ({\sqrt {g_{oo}}} g^{i\lambda}g^{k\nu} f_{\lambda\nu}).
%\cr \partial_o {\bm B}_{km} \equiv curl [({\bm j}/j^o) \times {\bm B}_{km}  ], \  {\bm E}_{km} \cdot {\bm B}_{km} \equiv 0, 
%\ {\bm E}_m \cdot {\bm j} \equiv 0.
 }
%\end {cases} 
\end{eqnarray}

The geometric 3-densities $j_{km}^\nu f_{\nu i}\varphi_{\!_G} \equiv (g_o j^oE^{km}_i   + [{\bm j}\times {{\bm B}_{km}}]_i)/\varphi_{\!_G} = 0  $ are similar to the specific Lorentz force with electric and magnetic fields. From the mechanical relations (4) it can be concluded  that the Maxwell-Lorentz equations for the moving densities of the extended electric charge $q_e$ are  local identities (not equations) for the correlated holism of electromagnetic fields and currents within the 3-volume integral of non-local electric energy $q_e \varphi_{\!_G} = const$. Likewise, electromagnetic balances for immeasurable microscopic currents and their fields should remain behind local identities of the Maxwell-Lorentz electrodynamics, as in the kinetic-metric balances (4) for the inertial continuum of non-local mass-energy.

\subsection {Adaptive geodesics  and kinetic pan-unity
	of the nonlocal cosmos}

Correlated changes of four kinetic potentials in the microscopic geodesics locally induce four metric potentials for both microscopics and for macroscopics in our mathematical approach to the monistic reality of positive energies. Thus, the microscopic kinetics initiate the correlated geometrization of matterspace-time  in the form of a non-stationary replica on the kinetic system self-governance, $ g_oj_{km}^o {\bm E}_{km}  +   {\bm j}_{km} \times {\bm B}_{km}  \equiv {\bm E}_{km} g_o j_{km}^o + {\bm B}_{km}  \times (- {\bm j}_{km})  = 0$ or 
\begin {eqnarray}
0 = \frac { c\varphi_{\!_G} } {4\pi { {g_o}}}  [ 
 {\bm \partial} (-\gamma g_o)  - \partial_o  (\gamma{\bm \beta}\!- \!\gamma{\bm g}) ]      {\bm {\partial}}\! 
\left [\! 
		 { (\!{\bm g}^2\!-1) {\bm E}_{km} \! -\! ( {\bm g}{\bm E}_{km}) {\bm g}   
		}  \! -\!{\bm g}\!\times\! {\bm B}_{km} \!\right ]\! + \cr \!
\!\frac {   
c{\bm B}_{km}} {4\pi {{g_o}}}\!\!
 \times \! \!
 \left [curl g_o ({\bm B}_{km}\! +\! {\bm g}\times {\bm E}_{km})\!
- \!\partial_o \!
\left (\! 
		 {  {\bm E}_{km} (\!1\!-\!{\bm g}^2\!)\! +\! ( {\bm g}{\bm E}_{km}) {\bm g}   
		}  \! +\!{\bm g}\!\times\! {\bm B}_{km} \!\right)\! 
		 \right],
			\end {eqnarray}
			in the non-local ethereal medium with constant mass-energy, momentum and angular momentum  integrals.
			
	Different probes with observable macroscopic bodies can test the geometrical self-assembly (4)-(5) from fast microscopic oscillations in the non-local matterspace.  Externally controlled probe bodies may start their free trajectories in the macroscopic geodesics (2) not with the equilibrium   3-velocity $v_i = cg_i$,
	  but with arbitrary initial values, when $cu_i \neq 0$ .  For instance, the steady rotation of the Earth material field in the meridian plane can maintain the equilibrium velocity $v_i$ of its inertial densities regarding the kinetically induced gyro-potential $g_i$,   $c u_i \equiv (cg_i - v_i)\gamma = 0$, while geodetic spacecrafts can move along any Kepler orbit and even in the retrograde direction, where ${\bm v}_{probe} \neq {\bm g}c$.
		The  microscopic counter-balance with the equilibrium metric potential,   when $g_i = v_i/c \equiv \beta_i, u^i \equiv c dx^i /ds \equiv \gamma v^i = 
\gamma cg^i,  u_o \equiv \gamma g_o =  g_o / {\sqrt {1-{\bm g}^2}} = const, u^o \equiv (1-{\bm g}{\bm v})\gamma/g_o = 1/g_o \gamma = 1/u_o $, can nullify both kinetic-metric strengths, 
${\bm E}_{km}\equiv - \varphi_{\!_G} [\partial_o \gamma({\bm \beta }  - {\bm g}) +{\bm \partial} (g_o\gamma)]/g_o 
 = 0$ and ${\bm B}_{km} \equiv  \varphi_{\!_G} 
curl ({\bm \beta }  - {\bm g} ) \gamma = 0$. Such an equilibrium in all space-time coordinates  corresponds to the potential self-organization of microscopic oscillations in the absence of external (non-metric) forces.

 The kinetic-metric inductions, 
${\bm D}_{km} \equiv  {\bm E}_{km} + {\bm g}\times\! {\bm H}_{km}   \equiv 
(1-{\bm g}^2){\bm E}_{km}  + ({\bm g}{\bm E}_{mk} ){\bm g} + {\bm g}\!\times\! {\bm B}_{km}= 0$,
 ${\bm H}_{km} \equiv  {\bm B}_{km} + {\bm g}\!\times \!{\bm E}_{km}   = 0$, are also nullified for the potential equilibrium of the microscopic background with ${\bm E}_{km} = {\bm B}_{km}=0$. 
From here, the potential four-flows in    
(4) exhibit the complete compensation of their current components, $\varphi_{\!_G}{\bm j}_{km} = 0$ and $\varphi_{\!_G} g_oj_{km}^o = 0$, for the Lomonosov-Umov energy transport  in kinetically induced metric potentials.
Such geometric self-assembly of invisible kinetic densities preserves the non-local correlations due to the kinetic-metric (locally adaptive) time rate until non-metric (external) influences reshape the microscopic geometrization of the distributed mass-energy. Again, there are no completely isolated systems in a line of the G$\rm{\ddot o}$del\rq{s} incompleteness theorems.

The inertia of massive densities during their geodesic acceleration is divided between the inward (ethereal) energy-chaos of Lagrange-Umov and the kinematic energy-order, called \lq Leibniz living forces\rq{} in Hamilton\rq{s} formalism. Unlike time-averaged microscopic densities, the probe bodies are too dense to match the equilibrium with weak metric fields in the macroscopic practice. The probe body tends to fall towards extremely dense kinetic regions of matterspace  and elastically bounces off them \cite{BulA}. When the motion occurs due to an external (non-metric) force in Newtonian mechanics, one can speak indeed about the 2nd law for the coerced  motion,  and not about the geodesic conservation of mechanical energy.

\section {Kinetic creation of  non-local mass-energy and measurable metric gradients}

\subsection {Correlated inertial fields and averaged metric replicas}

The kinetic self-organization of the inertial 3D continuum with the local time rate $d\tau \equiv g_\mu dx^\mu/c$ and correlated metric stresses can be tested by measurable accelerations of the probe (macroscopic, dense) body. Probe energies $m_pc^2$ in a steady material distribution are non-equilibrium interventions that fall on the Earth\rq{s} surface due to local Lomonosov pushes, and not due to long-range attraction from the underground or from the center of the planet.
  All observers assert that the moving centers of  visible bodies obey the 1914 Einstein relations (2) with metric potentials $<c^2g_\mu(x)>$ (or $c^2g_\mu (x)$, for simplicity) averaged over the macroscopic probe-body coordinates and the macroscopic time of probe-measurements. 
	
	In the generally accepted interpretations of macroscopic measurements, the original kinetics of microscopic oscillations  ware replaced by their phenomenological metric replica. And the  inverse square force ($m_pc^2{ \partial_r} g_o \approx - Gm_pM / {r}^2 $) has traditionally been associated with the fictional (yin) metrics of non-existing gravity, and not with kinetic fields. In our monistic mechanics  of only positive energies, the microscopic motion  in (1) induces both the repulsive (kinetic) potential $\gamma(r)$ and the attractive (metric) potential $g_\mu(r)$ of the correlated matterspace. Local probes of this non-local distribution with fast micro-oscillations  can  in practice reveal the non-equilibrium accelerations and velocities 	of very dense (visible) probe bodies according to  the macroscopic law (2).  This 1914 finding is not applicable to very low or invisible probe densities (comparable to or less than the energy density of the external metric field).

A freely falling observer in a moving elevator cannot define Lomonosov pushes in local measurements  due to the geodesic balance of metric stresses and acceleration responses of the probe body in  (2), where $dx^i \neq 0$.
Such a strict condition  implies that space-time derivatives of 
 the geometric four-potential, $ \partial_\nu   g_\mu $,  can only be determined by non-geodesic observers through measurable changes in  3-velocities $c\beta_i = \delta_{ij} c \beta^j \equiv c \delta_{ij} dx^j/g_\mu dx^\mu$ of the probe bodies.

 A non-geodesic observer, supported by external forces in the fixed system of reference with 
$dx^i = 0, <\gamma v^i> = 0$,  can measure two    
averaged  metric stresses, $ {\bm E}_{m}  \equiv    -  <\varphi_{\!_G} 
({\bm \partial} g_o - \partial_o {\bm g} )/g_o>\ \neq 0$ and ${\bm B}_{m}  \equiv -\varphi_{\!_G}  curl  < {\bm g} >\ \neq 0,
$ induced by fast oscillations of local ethereal densities. The time-averaged microscopic fields are also averaged over a dense volume of the macroscopic charge ${\sqrt G}m_p$ (to use the point center of inertia in Einstein\rq{s} geodesics). 
The non-local Lomonosov\rq{s}  liquid  always exerts only local metric pushes on the probe macro-densities  in contrast to the \lq action-at-a-distance\rq{} of Newton and further generalizations of dual physics.

A motionless observer on the ground feels stresses-pushes induced locally by the non-locally organized densities of the Earth, but not by the Sun (or the Galaxy, etc). This observer moves freely with vanishing kinetic-metric stresses in the Sun\rq{s} fields according to  Einstein\rq{s} relations (2). But the absence of net stresses from the Sun in the Earth\rq{s} laboratory does not mean the absence of  macroscopic Sun\rq{s} densities arising from microscopic oscillations beyond the Sun\rq{s} visible surface.  Zero local observations in the geodesic laboratory cannot deny the existence of material densities and their metric stresses  which geodesically accelerate this laboratory in the material cosmos. By denying the continuous matterspace of inertial energies from the available measurements, modern physics could similarly deny the existence of the Sun, assuming for a moment complete screening  of solar radiation and associated reflections from the planets.  

The fundamental balances (5) for  microscopic geodesics within the volumetric mass-energy integral
$M_{Sun}c^2 = const$ are consistent with the words of Plato: \lq\lq Matter and space are the same\rq\rq. 
And Einstein\rq{s} question \lq\lq Does that mean the Moon is not there when I am not looking at it\rq\rq{} can be answered that the Moon\rq{s} inertial macro-densities  and there metric stresses are always \lq there and here\rq{.} Nevertheless, without wave dissipation or other inelastic interventions into the metric-kinetic balance, a local observer cannot detect the Moon\rq{s} material extension according to the macroscopic law (2) for the elastic motion.

\subsection {Method of kinetic potentials to create measurable mass-energy }

The kinetically based geometrization of microscopic oscillations  can explain the adaptive responses of the covariant four-velocity $cu_\mu$ to  external (non-metric) forces in (1).  The non-stationary responses  have the high-order temporal derivatives of 3-velocities, which, after time-averaging, can quantitatively explain macroscopic flutter, maelstroms, and other turbulent phenomena of the non-local matterspace with feedback processes  \cite {BulTur}. How can the microscopic geodesics (5)  explain static, stationary or pulsating self-organizations of macroscopic systems? 		

For simplicity,  we analyze  only potential micro-circulations, ${\bm B}_{km} = {\bm E}_{km} = 0$, resulting in the static self-assembly of macroscopic densities around a center of spherical symmetry. In other words, the static equilibrium of macroscopic  densities can correspond
to stationary micro-oscillations of 3-velocities under the fixed four-potential, $ u_i = 0$ and $u_o = 1$. 
The stationary geodesics (5) give the time-averaged relations for macroscopic applications: $<\!cu_i\!>_o = <\!cg_i\gamma\!>_o - <\!v_i\gamma\!>_o = 0,  
<\partial_o (cg_i\gamma - v_i\gamma)\! >_o = 0, %{\bm E}_{km} \equiv -\varphi_{\!_G} [{\bm \partial} (\gamma g_o)]/g_o     , 
<\!{\bm B}_{km}\!>_o \equiv \varphi_{\!_G} curl (<\!{\bm v}\gamma\!>_o - <\!{\bm g}\gamma\!>_o )/c  = 0 $.  The potential microscopic states with radial auto-pulsations of $v_r(x^o) = cg_r(x^o) \neq 0  $ induce static macro-organization with 
$<\!{\bm v}\!>_o = <\!c{\bm g}\!>_o = 0 $,  $<{\bm v}^2>_o \equiv   
{\bm v}^2(r) =  c^2{\bm g}^2(r) \equiv <c^2{\bm g}^2>_o \neq 0,$ and   the local counterbalance of averaged metric $<{\bm e}_m>_o \equiv {\bm e}_m (r) \neq 0$ and  kinetic $<{\bm e}_k>_o \equiv {\bm e}_k (r) \neq 0$ stresses:
  \begin {eqnarray}
 \cases {  
  %<{{\bm E}_{km}}  div (-{\bm E}_{km} - {\bm g }\times [{\bm g }\times {\bm E}_{km}]) >  =
 %[{\bm \partial} (-\gamma g_o) ]      {\bm {\partial}}\! \left [\! 
	%	 { (\!{\bm g}^2\!-1) {\bm E}_{km} \! -\! ( {\bm g}{\bm E}_{km}) {\bm g}   
		%}  \right ], 
		%		({\bm e}_k + {\bm e}_m) div (- {\bm e}_k - {\bm e}_m  ) = 0,\cr
		<{\bm E}_{km}> = - <{\varphi_{\!_G}} \gamma {\bm {\partial}} ln [\gamma g_o]> =  {\bm e}_k(r)  +  {\bm e}_m(r) =0, 
		%\cr 		<{\bm v }> = c <{\bm g}> = {\hat {\bm r}} v(r)  = c {\hat {\bm r}} g(r) = 0,
		\cr <{\bm e}_k> 	
		\equiv - <{\varphi_{\!_G}}{\bm \partial} \gamma>  = - {\varphi_{\!_G}}{\hat {\bm r}}\partial_r \gamma(r) \equiv {\bm e}_k(r), 	
			\cr 	
		<{\bm e}_m > \equiv  -<{\varphi_{\!_G}}\gamma {\bm \partial} ln g_o> = -{\varphi_{\!_G}}{\hat {\bm r}}\gamma(r)
		\partial_r  ln g_o (r) \equiv {\bm e}_m(r), 
		\cr
		\gamma(r) \equiv 
		\frac {c}{\sqrt {c^2 - {\bm v}^2(r) }} = \frac {1}{\sqrt {1- {\bm g}^2(r) }}, \gamma(r) g_o(r) =  \gamma(\infty) g_o(\infty) = const \ (\Rightarrow 1). 
				}
 \end {eqnarray}

 The radial density $\varphi_{\!_G}\rho(r) \equiv - \varphi_{\!_G} div {\bm e}_m (r)/4\pi =  {\varphi^2_{\!_G} r^2_o}/{4\pi r^2(r+r_o)^2}$ of the spherical mass-energy integral $Mc^2 \equiv r_o  \varphi^2_{\!_G}$ 
 has already been found by the strong negative field ${e}_m (r) \equiv - {\sqrt G} M/[r(r + r_o)] \leq 0 $  in the flat-space reading \cite {Bul} of Einstein\rq{s} physics for the probe charge ${\sqrt G}m$. The monistic requirements of field mechanics reject  gravitational fields with negative energies and declare the purely kinetic origin of inertial bodies. Therefore, positive kinetic energies should derive the same material density $\varphi_{\!_G}\rho(r)$ through the positive kinetic stress ${e}_k (r) = -{\varphi_{\!_G}}\partial_r \gamma(r) \geq 0$ or negative gradients of a positive Poincar{\'e} pressure. 

The microscopic oscillation-fluctuations with $<{\bm v}^2> \neq 0$  produce the inhomogeneous kinetic potential $\varphi_{\!_G}\gamma(r) \geq c^2 / {\sqrt G}$, the static metric consequences $g_{oo}(r) = const/\gamma^2(r) \Rightarrow 1/\gamma^2(r) $ and $<{\bm g}> = 0$, and the missing  counter-pressure of Poincar{\'e} to balance the metric self-force in the static distribution of  macroscopic densities and fields. In this methodical way, the microscopic kinetics with geodesic  velocities generate the macroscopic counter-balance $g_o(r)\gamma(r)  = const$ for the kinetic and metric macroscopic  self-potentials of the non-local matterspace. The macroscopic metric tensor $g_{\mu\nu}(r)$, with $g^2_o(r) \equiv g_{oo}(r) \equiv <g_{oo}> \approx 1 - (2GM/rc^2), g_{oi}(r) = <g_{oi}> = 0, g_{ij}(r) \equiv <g_{ij}> = - \delta_{ij}$, originates from the eternally fluctuating-oscillating  micro-velocities in (2). The averaged relations for the static equilibrium of macro-densities and macro-stresses submit the following non-local solutions:      
\begin {eqnarray}
 \cases {  
		\varphi_{\!_G} \rho (r) \equiv		 \varphi_{\!_G} div {\bm e}_k (r)/4\pi  =  [{\bm e}^2_k(r) + {\bm e}^2_m (r)]/8\pi  \equiv  
		\	{\rm {average\ energy}} \cr 
		{ \rm { density\ of \ kinetic \ and  \ metric \ counter - oscillations}},
		\cr
		{\bm e}^2_k (r) \equiv (-\varphi_{\!_G}\partial_r \gamma)^2 = (-\varphi_{\!_G}\gamma \partial_r ln g_o )^2  
		\equiv {\bm e}^2_m(r) \ {\rm in \ stationary \ equilibrium},
		\cr
		  \varphi_{\!_G} \partial_r (r^2 e_k) = r^2 e^2_k, e_k+e_m = - \varphi_{\!_G}\gamma \partial_r ln (\gamma g_o) = 0,\   \gamma(r) = 1/ g_o(r), 
	\cr 	{\bm e}_k(r) =  {\hat {\bm r}} \frac { \varphi_{\!_G} r_o} {r (r+r_o)}  = {\hat {\bm r}} \varphi_{\!_G} 
	\left ( \frac {1}{r} -\frac {1}{r+r_o} \right)
	 = - {\bm e}_m(r) =  {\hat {\bm r}}\frac {\varphi_{\!_G}}{g_o(r)} \partial_r ln g_o(r),
	\cr  \frac {1}{r+r_o} -\frac {1}{r}  = \partial_r \left (\frac {1}{g_o} \right ), \ g_o(r) = \frac {1}{1+ 
	ln [1 +(r_o/r)]},
\	\gamma(r) = 1+ ln [1 +\frac  {r_o}{r} ], 
\cr 	 g^2(r) = v^2(r)/ c^2 = 1 - [ 1+ ln(1 + \frac {r_o} {r}) ]^{-2} = 1 - [1/\gamma (r)]^2, \ r_o \equiv \frac {GM}{c^2},
	\cr \rho_{eq}(r) = \frac {1}{4\pi}  {\bm \partial }{\bm e}_k (r) = - \frac {\varphi_{\!_G}}{4\pi} {\bm \partial }^2 \gamma (r)
											 = - \frac {\varphi_{\!_G}} {4\pi r^2}  \partial_r  (r^2 \partial_r \gamma ) =  
											\frac {\varphi_{\!_G} r^2_o}{4\pi r^2(r+r_o)^2}  .
}
 \end {eqnarray}

Again, a volume integral of the kinetic entity (an extended mass-energy $Mc^2$ with the static radial densities after the  macroscopic averaging) originates  from the correlated oscillations (fluctuations, vibrations, pulsations, etc.)  of microscopic velocities resulting in 
$v^2(r) = g^2(r) = 1- \gamma^{-2}(r) \neq 0$ and $\gamma(r) = 1+ ln [(r+r_o)/r] = 1/g_o (r)$. Elastic oscillations with radial falls and takeoffs are inherent not only to microscopic velocities  in  (5), but also to the macroscopic rebound  \cite {BulA} of probe masses in the  Einstein\rq{s} geodesics (2). 

\section {Quantum fields of correlated inertia }

\subsection {Kinetic nature of Poincar{\'e} counter-pressure}

The  microscopic oscillations with $<{\bm v}> = <{\bm g}> = 0$ do not generate 
the measurable spin (or gyro-moment) in such a nonlocal monopole with $ <{\bm g}^2> \neq  0$.  Radial auto-pulsations in (5) can be associated with Schr$\rm \ddot o$dinger\rq{s} zitterbewegung of the quantum particle \cite {Sch,Zhi}. For macroscopic physics, these micro-oscillations generate  radial gradients, $-\partial_r P_k(r)  \equiv  \rho_{eq} {\bm e}_k $ $
=  {\varphi^2_{\!_G}} (\partial_r \gamma )  \partial_r [   (\partial_r r^2 \gamma)/r^2 ]/4\pi
 > 0 $, of the inward kinetic pressure $P_k(r)$ in order to prevent the gravitational collapse of equilibrium densities 
$\rho_{eq}(r) = Mc^2 r_o / 4{\varphi_{\!_G}}\pi r^2 (r+r_o)^2$ under the self-action from emerging metric fields $e_m(r) < 0 $ within the non-local distribution of $Mc^2$:  
\begin{eqnarray}
\cases {
 -{\bm \partial} P_k (r)  \equiv \rho_{eq} {\bm e}_k   = - \rho_{eq} {\bm e}_m  = 
{\varphi_{\!_G}}\rho (r)\gamma(r) {\bm \partial} ln g_o(r) = 
 +\frac { {\hat {\bm r}} \varphi^2_{\!_G} r^3_o} {4\pi r^3(r+r_o)^3}
\cr 
P_k(r) - P_k(\infty) = - \frac { \varphi^2_{\!_G} }{4\pi r^3_o} \int\left ( \frac {r_o}{r} - \frac {r_o}{r+r_o} \right )^3dr 
\cr = 
 \frac { Mc^2 }{4\pi }\left [ 
\frac {1}{2r_o r^2}  - \frac {1}{2r_o (r+r_o)^2}
+ \frac{6}{r^3_o} ln {\left (1+ \frac {r_o}{r} 
 \right )} - \frac {3}{r^2_or} - \frac {3}{r^2_o(r+r_o)}\right ]
\approx \frac { Mc^2 r^2_o}{20\pi r^5 }
 > 0.
%\rho {\bm e}_k \equiv -{\hat {\bm r}} {\varphi_{\!_G}}\rho (r)\partial_r \gamma(r) 
%\equiv + \frac {{\hat {\bm r}} \varphi^2_{\!_G} r^3_o}{4\pi r^3(r+r_o)^3} \cr
%=
%-\rho {\bm e}_m  \equiv {\varphi_{\!_G}}\rho (r)\gamma(r) {\bm \partial} ln g_o(r) \equiv 
%+\frac {{\hat {\bm r}} \varphi^2_{\!_G} r^3_o}{4\pi r^3(r+r_o)^3}.    
}\end {eqnarray}

Equilibrium mass density  ${\varphi_{\!_G}}\rho_{eq}(R)/c^2 = M_\oplus   r_\oplus/{4\pi R^2(R+r_\oplus)^2} \approx 1.3 \times 10^{-9} g/cm^3 $ of the Earth\rq{s} invisible extension in the middle, e.g. of the stratosphere ($h = 30 km$, $R_\oplus= 6 371 km,   R =R_\oplus +h = 6, 401\times 10^8 cm, M_\oplus = 5,972\times 10^{27} g, r_\oplus =0,4435 cm $),  about nine orders of magnitude smaller than the mass densities of the probe bodies. Poincar{\'e} kinetic pressure at the same 30 km above the Earth\rq{s} surface, $P(R_\oplus +h) \approx (5.9\cdot 9 \times 10^{40} J/20\pi) \cdot (0.0044 m)^2/ (6.4 \times 10^6 m)^5 = 15 \ Pa = 1.5 \times 10^{-4} \ bar , $ cannot provide the Archimedean force to repel meteorites and other heavy bodies from the stratosphere. Here, the 1914 geodesics (or the Lomonosov local push toward the Earth\rq{s} surface) dominate for the  probe densities of $5 g/cm^3$ $ \gg \rho_{eq} (R_\oplus +h) \approx 1.3 \times 10^{-9} g/cm^3$ .

Again, the mass density ${\varphi_{\!_G}}\rho_p(r)/c^2 \equiv \mu^\prime_p =const$ of a probe body in central fields is to be  compared with $\rho_{eq}(r)/ {\sqrt G}$ balanced by the kinetic pressure $ P_k(r)$ for spherical self-distribution of the mass integral $M \equiv r_o c^2/G$. For $\rho_p/{\sqrt G} \gg \rho_{eq}(r)/ {\sqrt G}$, the Einstein geodesic of 1914 works, and the probe mass moves with negative radial acceleration with relatively small radiation losses. For $\rho_p/{\sqrt G} \approx \rho_{eq}(r)/{\sqrt G}$
the probe densities are not constant, but change adaptively along the trajectory due to significant wave exchange and  dissipation. The latter goes beyond the  original assumptions of  the 1914 elastic geodesics.

The Poincare pressure $P_k(r)$, introduced quantitatively in (8) by the scalar  kinetic potential $ \gamma(r) \equiv <\gamma>\neq const$, when $<\gamma v_i/c> \equiv 0$,   diverges in the strong-field region, where  $rc^2/GM \ll 1$. In this region, the microscopic geodesics should be described by spinning solutions of (5) with non-zero kinetic, $<v_i\gamma/c> \neq 0 $, and metric $<g_i\gamma> \neq 0 $ gyro-potentials. The macroscopic equilibrium  with constant spin and mass integrals can provide a more accurate  approximation for the resulting metric fields and mass-energy densities from the averaged relations (5). For example, one should derive $g_i (x,y,z) \neq const $ for the differential rotation  of  material densities in a spiral galaxy and for the  inevitable rotation during accretion to the center of a strong-field organization.

In general, the monistic ideas of Cartesian metaphysics and  Russian Cosmism for microscopic (etheric) oscillations and the corresponding distributions of macroscopic inertia lead to verifiable mathematical laws for kinetic and geometric counter-stresses with positive mass-energies in an isolated system.
For such balanced stresses, the kinetic origin of inertial densities and their bulk masses corresponds to the stationary Kant cosmology and explains the Bentley\rq{s} paradox in non-local self-organizations through the Poincar{\'e} kinetic pressure in the emerging masses.

\subsection {Quantization of inertial fields and volumetric integrals of  non-local mass-energy}

According to the primacy of the kinetic stresses ${\bm e}_k$  in the microscopic geodesics (5) and their radial solutions (6)-(7), the measurable metric field ${\bm e}_m(r) = - {\bm e}_k(r)$  is not an independent energy concept, but an observable consequence of the adaptive self-organization of the kinetic densities in their non-local cosmos. Therefore, macro-gravity is not a fundamental notion, but only a secondary replica of micro-kinetic correlations in the isolated system.
In such a holistic approach to all material densities and their kinetic stresses, it seems appropriate to start studies of quantum properties from primary (kinetic, yang) energies, rather than from macroscopic (metric, yin) replicas averaged over the world parameter $x^o \equiv c t$.

At first it was thought  that the main field variable itself
(Einstein\rq{s} metrics, not micro-oscillations) determines the structure of space-time, which determines the evolution of material densities. But the quantization of Einstein\rq{s} metric components $ <g_{\mu\nu}>_o$ for the macroscopic gravity could not lead to a consistent theory of gravitational waves (tensor gravitons) with quantized  energy-momentum. Already after the first attempts \cite {Bro, Sny, Gup, DeW} to quantize a non-existent geometric entity with negative gravitational energy and momentum, the internal inconsistency of the theory \lq\lq Quantum Gravity\rq\rq{} began to be discussed in many advanced studies \cite {Lat, Rov, Kie, Bor}.

The declared conceptual difficulties in the supposed   quantization of gravitational fields, regardless of their material sources or kinetic energies, may suggest that the metric relativity, in our monistic view, return to the Cartesian mainstream of pure field matter. Even before the quantum mechanics of wave matter, Gustav Mie  \cite {Mie} described the extended electron as a  bundle  of field energy (without a point source) through the proposed equation, $\nabla^2 A_o + 4\pi g^2 A^5_o =0,$ for the static potential $A_o$ of Maxwell\rq{s} theory. Then the found solution $A_o (r) = q/ (r^2+r^2_g)^{1/2}$, with 
$r^2_g \equiv 4\pi g^2 q^4/3$, turned out to be unstable for a radial electron. But the idea of using Maxwell\rq{s} physics for the field theory of continuous matter was generally correct. From the present considerations we can say that behind the micro-kinetic mechanism of inertial mass and the measurable gradients of $g_{o\mu}$ are Maxwell-type relations (4)-(5) for local geodesic balances.

In the monistic background of microscopic oscillations, the system mass-energy should consist of elementary wave energies $\hbar \sum_1^{_N} \omega_n(t) \equiv \sum_1^{_N} m_n (t)c^2 = M_{sys} c^2 = const $. The fundamental equality of elementary masses and frequencies in  quantum mechanics, $ \hbar \omega_n (t)\equiv m_n(t)c^2  \Rightarrow r_n(t) \varphi^2_{\!_G}$, can be represented for the radial organization (7) of an isolated mass-energy 
($m_1 c^2 = const$) by the spatial integral of its kinetic field ${\bm e}_k(r,t)  =  - {\bm \nabla }\gamma (r,t) \neq 0$, when $\partial_o \gamma ({\bm \beta - \bm g}) = 0 $,
\begin {eqnarray}
\cases {  
	{\bm e}_k(r,t)  = {\hat {\bm r}} \frac { m_n[t] c^2} { \varphi_{\!_G} r (r+r_1[t])}  = {\hat {\bm r}}
	  \frac {\varphi_{\!_G}}{r}
	\left ( 1 -\frac { 1}{1 +   (\hbar \omega_1[t]/r \varphi^2_{\!_G})} \right),
	\cr
	\int {\bm e}^2_k(r, t)\frac { dV}{4\pi} = 
	{\varphi^2_{\!_G} } \int_o^\infty  \left (\frac { \hbar \omega_1 [t]}{r \varphi^2_{\!_G}\ + \ 
		 \hbar \omega_1 [t]} \right)^2 dr = \hbar \omega_1(t) = m_1 c^2 = const, 
	\cr
	\gamma^{-2} (r) = 1- \beta^2(r)   = [1 + ln (1 +(  \hbar \omega_1/r\varphi^2_{\!_G}  )  ]^{-2} = g_{oo} (r) .
}
\end {eqnarray}

If the continuous matterspace has two or more peaks of energy densities \cite {BulCol}, the  wave pattern loses its  spherical symmetry due to mutual correlations of elementary densities in the non-local system of continuously distributed inertia with a constant integral $M_{sys}c^2$:
\begin {eqnarray}
\cases {  
		{\bm e}_k({\bm x}, t)  =  - {\bm \nabla }\gamma (r,t) =  \frac {\sum  m_n[t] c^2  \frac { ({\bm x} - {\bm a}_n[t])  }{|{\bm x} - {\bm a}_n[t]|^3 }} { \varphi_{\!_G}  [1+ \sum (r_n/|{\bm x} - {\bm a}_n [t]|)] }
		 = \frac {\varphi_{\!_G} \sum r_n [t]  \frac { ({\bm x} - {\bm a}_n[t])  }{|{\bm x} - {\bm a}_n[t]|^3 }} {  [1+ \sum (r_n[t]/|{\bm x} - {\bm a}_n[t]|)] },
\cr\cr
\partial_o \gamma { \bm g} = \partial_o \gamma {\bm \beta }, \
{\bm e}^2_k({\bm x},t) \equiv { \varphi_{\!_G}   }div {\bm e}_k ({\bm x},t), 
\cr\cr
	 M_{sys}c^2 \equiv  \int {\bm e}^2_k({\bm x}, t)\frac { d^3x}{4\pi}
	 \equiv \int \frac { d^3x}{4\pi\varphi^2_{\!_G} }\frac { 
	 	\left (\sum \hbar \omega_n [t]  \frac { ({\bm x} - {\bm a}_n[t] )  }{|{\bm x} - {\bm a}_n[t]|^3 } \right )^2 } {  [1+ \sum \hbar \omega_n[t]/(\varphi^2_{\!_G}|{\bm x} - {\bm a}_n[t]|)]^2 }
 	\cr 
 	\equiv  \frac { \varphi_{\!_G}} {4\pi }\! \int\!  { d^3x} div  {\bm e}_k ({\bm x}, t)\equiv  \frac { \varphi_{\!_G}}{4\pi }\!\int\! {\bm e}_k d{\bm S}_{_R\rightarrow \infty}    = \sum \hbar \omega_n(t) = const ,
	\cr\cr
	\gamma (\bm x,t) =  [1 + ln (1 + \sum (  \hbar \omega_n[t]/|{\bm x} - {\bm a}_n[t]|\varphi^2_{\!_G}  )  ] = 1/g_{o} (\bm x,t) .
}
\end {eqnarray}
Here we used the field equivalences $div [({\bm x} - {\bm a}_n[t]) /|{\bm x} - {\bm a}_n[t]|^3] \equiv 0$. 
They preserve the Einstein\rq{s} equivalence principle for heavy and inertial densities in the continuous matterspace, where ${\bm e}^2_k/ {4\pi} = { \varphi_{ \! _G} }  {\bm \nabla}{\bm e}_k  / {4\pi } $. 
The new quantization method  in terms of the correlated kinetic field with the  4D geometric organization allows us to quantize the field integral of the non-local mass-energy, $M_{sys}c^2 \equiv \sum \hbar \omega_n $. All time-varying peaks of elementary energies are assumed  to be in finite motion relative to the common center of inertia. 
Therefore,  a system scale $|{\bm a}_n (t)|_{ max}$ of the most significant densities  in (10) is smaller than the chosen  radius R of the Gaussian surface. For macroscopic scales of the coordinate time $t = x^o/c$, one can admit $\partial_t {\bm a}_n(t) \neq 0$ and $\partial_t \omega_n (t) \neq 0$ in (10) even for a steady structural self-organization of the isolated energy system with  $M_{sys}c^2 = const$.

In general, the Maxwellian analogy in microscopic counterbalances (4)-(5) makes it possible to build quantum regularities into the covariant relations of General Relativity, provided  that they are translated into the field language  of  wave packets in continuous mass - energy or material continuum-plenum of the ancient Greeks.
As is well known, Einstein believed that quantum mechanics is not a  complete theory and must follow from covariant laws for field forms of dense matter (called bunched fields). Then the quantization of the first integrals of motion should arise as additional integral conditions that close the differential set of local mathematical equations. This is the perspective of the ethereal (microscopic) approach  of Russian cosmists, where the non-local pan-unity of the entire hierarchy of material formations is subject to Umov\rq{s} local transfers with reversible mutual transformations of chaos energy and kinetic order.

\section {Discussion } %Conclusion

Dual Newton-Einstein physics was developed for macroscopic measurements, where gravitational or metric fields appear to control the accelerated motion of probe bodies. Monistic field mechanics and microscopic mathematics (3)-(8)  are concerned with the primary cause of measurable fields and the non-local laws of their correlated changes.
The kinetic organization of the macroscopic 3D-continuum and 4D-geometry with a local balance of kinetic (Poincaré) pressure and local metric replicas revises the Newtonian scheme of gravitational attraction between distant bodies in the empty space. 
The local self-control in (5) or the adaptive responses  of  microscopic oscillations in (1)  imply inevitable correlations on micro-, macro- and mega-scales in the global hierarchy of non-local organizations.

The microscopic oscillations in the monistic pan-unity of Russian and European cosmists obey the Hegelian dialectics and yin-yang relations that can  quantitatively describe the kinetic energy continuum, including inert and living subsystems in the common Geosphere, Bioshere, and  Noosphere of the Earth.  Here, Western dialectical thinking has focused mainly on local measurements or observations in external fields according to (2) and the pragmatic dominance of Newtonian mechanics for non-isolated bodies. But earlier Eastern teachings have traditionally focused on the holistic organization of field energies throughout the entire cosmos, described by the local geodesics (5). Field bifurcations  for the structural reorganization in (1) depend  on non-metric (inelastic) interventions in nearly elastic micro-oscillations in the continuous matterspace.

	The continuous distribution  of inertial densities in non-local matterspace is conceptually contradicts the Newtonian empty space for distant gravitation or the standard dualism of modern physics for separated particles and their  massless fields/forces. Therefore, Lomonosov\rq{s} local pushes and the ethereal school of Stoletov-Umov were not supported by the domestic Academy of Sciences. Nevertheless, Lomonosov\rq{s} ethereal physics can quantitatively explain Newton / Coulomb inverse squires in macroscopic practice  \cite {BulN, BulCol}. This old Russian cosmos can also shed some light on the observations of  non-locally correlated quasars \cite {Ali}, the mysterious  motion of coordinated galaxies \cite {Mys}, the unprecedented stability of the solar system and other cosmic formations. New geodetic identities  (3)-(8) can  quantitatively explain slowly pulsating and stable cosmic densities in Kant\rq{s} rational cosmology. 
	
	The emerging experimental evidence of entangled macro-megastates can be a clear reason to include the kinetic pressure (8) of inward micro-oscillations into the monistic world of mass-energy organizations and to revise the classical dualism of matter with independent kinetic and gravitational energies. Contrary to the concept of fictitious (negative, non-existent) gravitational energies, the positive kinetic energy in (7) can lead to a new mechanism of mass creation, can form a 4D pseudo-Riemann  geometry with the 3D Euclidean section, and can describe macroscopic phenomena in the monistic (kinetic) reality  with pure field material densities.  
					
	The 1873 reversible energy $mc^2/\gamma \leq mc^2$  of the Umov\rq{s} ether is an inward kinetic degree of freedom? which allows a monistic
	kinetics of pulsating falls and bounces \cite{BulA} without the need of additional gravity-concepts. The latter cannot  explain the Bentley\rq{s} paradox, Kant\rq{s} stable cosmology  and the Nietzsche \lq\lq eternal return of the same\rq\rq{} \cite{Nie} from  centers of strong fields. Newtonian empty-space gravitation cannot satisfactorily account for the steady galaxies, the observed astronomical jets, the probable auto-pulsations of the Metagalaxy, and the unusual behavior of many other cosmic objects \cite{Nat}.
		Our  developments of old ethereal ideas and new mathematical physics in terms of monistic energy from microscopic oscillations clarify the kinetic nature of Poincar{\'e} pressure and the appropriate quantum formalism for the metric theory of non-local inertia. 
		
		Mechanical waves of massive fields can coherently converge quantum and kinetic-metric approaches to continuous matterspace. Adaptive physical time is required for the self-consistent transport of inertial densities and  correlated (timeless) stresses in the isolated self-organization of their non-local energy integral. 	The current search for quantum gravity can be redirected to the relativistic  path of quantum inertia for purely kinetic energies with wave chaos and translational order. The mathematical unification of field mechanics,  electrodynamics, thermokinetics and magneto-hydrodynamics  in common terms of quantum and geometrical characteristics is still  one of the frontal problems of modern natural science.
	
%\bigskip \noindent {\bf		AUTHOR CONTRIBUTIONS}
%		{The author confirms being the sole contributor of this work and
%		approved it for publication.}		
						
%$\bigskip \noindent {\bf Conflict of interest statement} There is no conflict of interest to declare.

%\bigskip \noindent {\bf Acknowledgments} 

%\bigskip \noindent {\bf ACKNOWLEDGMENTS}

%The author acknowledges the useful discussion at the seminar for Time Nature Exploration named after Prof. A.P. Levich. %The author is also grateful to the editors and   reviewers for careful reading and professional comments.

%\bigskip \noindent {\bf Funding} 

%\bigskip \noindent {\bf FUNDING}

%Open Access funding enabled by the PSAL program  of
 %Peoples\rq{} Friendship University of Russia. 

%\end{multicols}{0}


\begin{thebibliography}{999}
%\bibliography {}
	
	
		
	\bibitem{Des}  Descartes\rq{} Physics, in { The Cambridge Companion to Descartes} (ed. Cottingham, J.), 
Cambridge University Press, New York, 1992, pp.286-334.


\bibitem{Poi} H. Poincar{\'e},  %Henri, "Sur la dynamique
%de l\lq {\'e}lectron (received on 23 July,1905), Rendiconti del Circolo Matematico di Palermo 21, 1906, pp.
%129-175 (1-47).
{ On the dynamics of the electron}, {Rendiconti del Circolo Matematico di Palermo} { 21} (1906) 129.

	
	\bibitem {BulT} I.{\'E}. 	Bulyzhenkov,   Thomson 4/3 problem leads to nonlocal continuous charges with Poincaré radial stresses and zero electromagnetic inertia, { Physics Letters} { A}  383 (2020) 2367-2369.

\bibitem {Lom}	M.V. Lomonosov,  Notes on the severity of bodies. In Complete Works, 11 Vols.; Eds. Vavilov, S., Kravetz, T., Eds.; Akad. Nauk. USSR, Moscow and Leningrad, USSR, 1950; Volume 2.

\bibitem{BulN} I.{\'E}. Bulyzhenkov,
	Monism of Nonlocal Matterspace with Instant All-Unity Instead of Particle–Field Duality with Retarded Interactions
Phys. Sci. Forum  7(1) (2023) 48. %https://doi.org/10.3390/ECU2023-14031
	


\bibitem {BulCol} I.{\'E}. 	Bulyzhenkov, 
Coulomb Force from Non-Local Self-Assembly of Multi-Peak
Densities in a Charged Space Continuum, Particles 6 (2023)
136–143. %https://doi.org/10.3390/particles6010007
	

\bibitem {Umo} N.A. Umov,  Beweg-Gleich. d. Energie in contin. Korpern. Schomilch. Zeitschriff d. Math. und Phys. 1874, XIX (In Collected works., Gostekhizdat, Moscow-Leningrad, USSR, 1950, in Russian).
	
	
	
	
	\bibitem{Mie}	G. Mie, Grundlagen einer theorie der materie, {Ann. der Physik} { 37} (1912) 511-534; { 39} (1912) 1-40; 
	{ 40} (1913) 1-66.

	
	\bibitem{Bul} I.{\'E}.	Bulyzhenkov,  Einstein\rq{s} gravitation for Machian relativism of nonlocal energy-charges, { Int. J. Theor. Phys. } { 47} (2008) 1261-1269.   %https://doi.org/10.1007/s10773-007-9559-z

	
	
	
	
		
	\bibitem{EI} A. Einstein, L.  Infeld, { The Evolution of Physics}, Cambridge Press, {Cambridge}, {1938}.
	
	
	




\bibitem{Lan}  L.D. Landau, E.M. Lifshitz, { The Classical Theory of Fields: Volume 2}, Course of Theoretical Physics Series, 4th Edition, Butterworth-Heinemann, 1980.

\bibitem {BulTur}I.{\'E}.  Bulyzhenkov, Metric inertia for eddy densities of nonlocal matter-space. Journal of Turbulence,  21 (2021)  623-639. % DOI: 10.1080/14685248.2021.1953698


	
	\bibitem {Nie} D.F. Ferrer,  { Nietzsche\rq{s} notebook of 1881: The Eternal Return of the Same}
 (Verden, Germany: Kuhn von Verden Verlag, 2021).

\bibitem {BulA} I.{\'E}.  Bulyzhenkov,   Gravitational attraction until relativistic equipartition of internal and translational kinetic energies. { Astrophysics and Space Science  } { 363} (2018)  39. 
	

\bibitem{BulTests} I.{\'E}. Bulyzhenkov, Geometrization of Radial Particles in Non-Empty Space Complies with Tests of General Relativity.  Journal of Modern Physics 3 (2012) 1465-1478.


%\bibitem{BulF} Bulyzhenkov, I.{\'E}. 
	% Densities of complex charges unify particles with fields and gravity with electricity. {\it Bulletin Lebedev Physics Inst. } {\bf 43},  138 (2016).







%\bibitem{BulG} I.{\'E}. Bulyzhenkov,  Cartesian material space with active-passive densities of complex charges and yin-yang compensation of energy integrals, {Galaxies  } { 6}(2) (2018), 60.%; https://doi.org/10.3390/galaxies6020060



\bibitem{Sch}
E. Schr$\rm {\ddot o}$dinger,   $\rm {\ddot U}$ber die kr$\rm{\ddot a}$ftefreie Bewegung in der relativistischen Quantenmechanik [On the free movement in relativistic quantum mechanics] (in German, 1930), Sitzungsberichte der Preuischen Akademie der Wissenschaften, Physikalisch-mathematische Klasse, 1 (1930),
 418–428. 
 
 

\bibitem{Zhi}
W.  Zhi-Yong and X. Cai-Dong, Zitterbewegung in quantum field theory. Chinese Physics B 17(11) (2008) 4170.


	
	
		
	\bibitem{Ali}  D. Hutsem{\'e}kers, L. Braibant, V.  Pelgrims, D.  Sluse,  Alignment of quasar polarizations with large-scale structures, 
{Astronomy and Astrophysics }  572 (2014)  { A18} .

\bibitem{Mys} J.H.  Lee, M. Pak, H. Song,  H.-R. Lee, S. Kim, H. Jeong,  Mysterious coherence in several-megaparsec scales between galaxy rotation and neighbor motion, { Astrophysical journal}  { 884}(2) (2019) 104.

		
	
	
		
	
\bibitem{Nat} N. Hurley-Walker, X. Zhang, A. Bahramian, S. J. McSweeney, T. N. O’Doherty, P. J. Hancock, J. S. Morgan, G. E. Anderson, G. H. Heald, and T. J. Galvin,  A radio transient with unusually slow periodic emission,
Nature  601 (2022) 526–530.

\bibitem{Bro} M. Bronstein, Quantentheorie schwacher Gravitationsfelder, Phys. Z. Sowjetunion 9 (1936) 140 - 157. 


\bibitem {Sny} H.S. Snyder, Quantized space-time, Physical Review 71 (1947) 38-41.

\bibitem{Gup}
S. Gupta, Quantization of the Gravitational Field, General Theory, Proc. Phys. Soc.  A65 (1952) 608–619.

\bibitem{DeW}
B.S. DeWitt, Quantum Theory of Gravity, II. The Manifestly Covariant Theory. III. Applications of the Covariant Theory, Phys. Rev. 162.  (1967) 1195–1233, 1239–1256.

%m{Wal}
%R. Wald, General Relativity, The University of Chicago Press, 1984.

\bibitem{Lat} B.N. Latosh, 
Basic problems of conservative approaches to a theory of quantum gravity, 
Physics of Particles and Nuclei 51 (2020) 859-878.




\bibitem{Rov} C. Rovelli, Quantum Gravity,  Cambridge, UK: Cambridge University Press, 2004, 455 p.

\bibitem{Kie} C. Kiefer, Quantum Gravity, Oxford: Oxford University Press, 2007, 355 p. 

\bibitem{Bor} H.-H. Borzeszkowski and H.J. Treder, H. J. The Meaning of Quantum Gravity. Springer Science \& Business Media, 2012. %ISBN 9789400938939.




	 
\end{thebibliography}
\end {document}